\documentclass[preprint,prb,floatfix]{revtex4}

\usepackage{graphicx}

\newcommand{\comment}[1]{}

\begin{document}

\title{Reply to "Sharp-mode coupling in High-Tc Superconductors"}

\author{Jungseek Hwang$^{1}$, Thomas Timusk$^{1}$, and Genda Gu$^{2}$}
\affiliation{$^{1}$Department of Physics and Astronomy, McMaster
University, Hamilton ON L8S 4M1, Canada\\ $^{2}$Department of
Physics, Brookhaven National Laboratory, Upton, New York
11973-5000, USA} \email{timusk@mcmaster.ca}

\begin{abstract}

{Our optical technique has the advantage of being a bulk probe,
which is less subject to uncertainties in the doping level and in the
quality of the surface than angle resolved photoemission (ARPES).
It is also capable of higher energy resolution and the overall noise
level is lower. The disadvantage is that it gives momentum
averaged properties. In light of these differences, it came as a
surprise to us that our reported optical
self-energies~\cite{hwang04} were able to track in accurate detail
the ARPES self-energies of Johnson {\it et al.}~\cite{johnson01}.
Our data indicate that, as a function of doping, not only could
both optical and ARPES techniques resolve the sharp mode from the
background but also that the sharp mode intensity decreases
uniformly, disappearing completely at a doping level of 0.23. As
superconductivity is still strong at this doping level, with a
T$_{c}$ of 55 K, we conclude that the sharp mode was not an
important contributor to high-temperature superconductivity.}

\end{abstract}


\maketitle

\comment{Introduction}

Cuk {\it et al.}~\cite{cuk04} make two points that optical data
may be insensitive to strongly momentum-dependent signals because
they are momentum-averaged, and also that in their ARPES
data~\cite{gromko03} for momenta near the antinodal point
($\pi$,0), the sharp resonance persists in the highly overdoped
region and does not disappear as we claim.

Although the measurements of Johnson {\it et al.}~\cite{johnson01}
were performed at the nodal point, the weakening of the resonance
also takes place at the antinodal point, as indicated by other
ARPES work~\cite{kim03}. As shown in Fig. 1, self-energy effects
at ($\pi$,0) are strongly doping dependent, joining the
normal-state background at a doping level of 0.24 --- just as they
do in our optical results and in the ARPES data of Johnson {\it et
al.}~\cite{johnson01} at the nodal point. All three experiments
show the same strong doping dependence.

It is therefore surprising that the work of Gromko {\it et
al.}~\cite{gromko03} fails to confirm these results. These authors
do not present doping-dependent plots of the self-energy, but a
visual inspection of Fig. 2 of ref.[4] suggests that the
self-energy effects are almost doping independent --- in contrast
to the marked doping dependence reported by Kim {\it et
al.}~\cite{kim03}. One reason for the disagreement may be the
difficulty in controlling the doping level in the surface layers
under the ultra-high-vacuum conditions used in the the ARPES
experiments.

\begin{figure}[t]
  \vspace*{0.0cm}%
  \centerline{\includegraphics[width=4in]{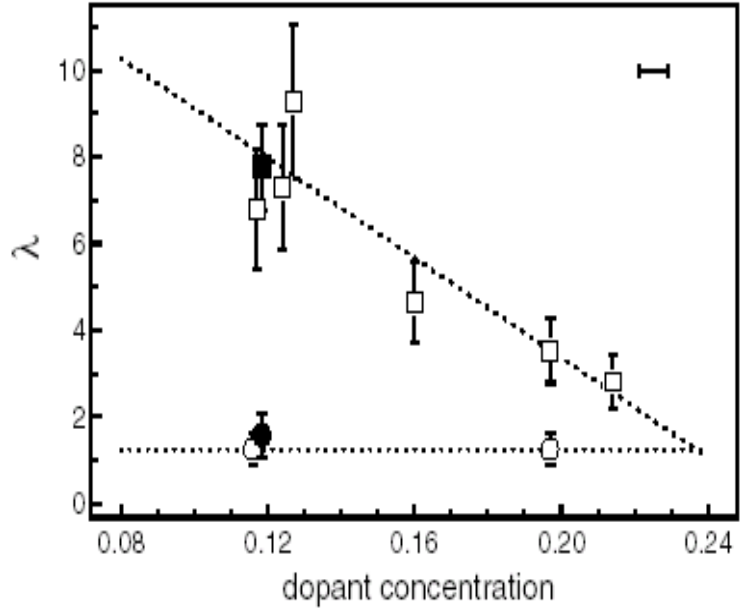}}%
  \vspace*{0.0cm}%
\caption{The coupling-strength parameter $\lambda$ as a function
of dopant concentration (see Fig. 3 from Ref.[5]). Squares,
superconducting state; circles, normal state; open symbols,
bonding band; filled symbols, antibonding band. Dashed lines are
straight-line fits to the data. Horizontal bar, experimental error
in the dopant concentration. Reprinted with permission from ref.
[5]; copyright(2003) of American Physical Society.}
\end{figure}

\end{document}